\documentclass[conference]{IEEEtran}
\IEEEoverridecommandlockouts
\usepackage{cite}
\usepackage{amsmath,amssymb,amsfonts}
\usepackage{algorithmic}
\usepackage{graphicx}
\usepackage{textcomp}
\usepackage{xcolor}
\usepackage{caption}
\usepackage{subcaption}
\usepackage{physics}
\def\BibTeX{{\rm B\kern-.05em{\sc i\kern-.025em b}\kern-.08em
    T\kern-.1667em\lower.7ex\hbox{E}\kern-.125emX}}

\begin{document}

\title{Trials and Tribulations of Developing Hybrid Quantum-Classical Microservices Systems}

\author{\IEEEauthorblockN{Javier Rojo}
\IEEEauthorblockA{\textit{Department of Computer and Telematic Systems Engineering} \\
\textit{University of Extremadura}\\
Cáceres. Spain \\
javirojo@unex.es}
\and
\IEEEauthorblockN{David Valencia}
\IEEEauthorblockA{\textit{Department of Computer and Telematic Systems Engineering} \\
\textit{University of Extremadura}\\
Cáceres. Spain \\
dvaleco@unex.es}
\and
\IEEEauthorblockN{Javier Berrocal}
\IEEEauthorblockA{\textit{Department of Computer and Telematic Systems Engineering} \\
\textit{University of Extremadura}\\
Cáceres. Spain \\
jberolm@unex.es}
\and
\IEEEauthorblockN{Enrique Moguel}
\IEEEauthorblockA{\textit{Department of Computer and Telematic Systems Engineering} \\
\textit{University of Extremadura}\\
Cáceres. Spain \\
enrique@unex.es}
\and
\IEEEauthorblockN{Jose García-Alonso}
\IEEEauthorblockA{\textit{Department of Computer and Telematic Systems Engineering} \\
\textit{University of Extremadura}\\
Cáceres. Spain \\
jgaralo@unex.es}
\and
\IEEEauthorblockN{Juan M. Murillo}
\IEEEauthorblockA{\textit{Department of Computer and Telematic Systems Engineering} \\
\textit{University of Extremadura}\\
Cáceres. Spain \\
juanmamu@unex.es}
}

\maketitle

\begin{abstract}
Quantum computing holds great promise to solve to problems where classical computers cannot reach. To the point where it already arouses the interest of both scientific and industrial communities. Thus, it is expected that hybrid systems will start to appear where quantum software interacts with classical systems. Such coexistence can be fostered by service computing. Unfortunately, the way in which quantum code can be offered as a service still misses out on many of the potential benefits of service computing. This paper takes the traveling salesman problem, and tackles the challenge of giving it an implementation in the form of a quantum microservice. Then it is used to detect which of the benefits of service computing are lost in the process. The conclusions help to measure the distance between the current state of technology and the state that would be desirable in order to have a real quantum service engineering.
\end{abstract}


\section{Introduction}

Quantum computing has been a relevant research field for more than 20 years, bringing together classical information theory, computer science, and quantum physics \cite{steane1998quantum}. More recently, the development of quantum computers has brought us to the Noisy Intermediate-Scale Quantum (NISQ) era \cite{preskill2018quantum}, where quantum computers with more than 50 qubits, although limited by the noise in quantum gates, are starting to perform tasks that may surpass the capabilities of classic computers.

Alongside this scientific development, quantum computing is also experiencing a significant commercial growth \cite{macquarrie2020emerging}. Several major computing corporations have built their own quantum computers and are offering them to users, mostly in a pay-per-use model. Engineers have designed and implemented dozens of quantum programming languages, simulators, and toolkits. All of this is paving the way for the development of quantum software and services.

Nevertheless, for the time being, classical and quantum services must not only coexist but interact with each other \cite{sodhi2018quality}. This coexistence has been called by some researchers hybrid classical-quantum systems \cite{mccaskey2018hybrid,mccaskey2020xacc}. A natural way to approach such collaborative coexistence is by following the principles of service engineering and service computing.  

Already, companies and researchers are leaning towards the use of quantum infrastructure and quantum software as a service, as they are used to do with classical computing resources. Offerings like IBM Quantum Computing\footnote{https://www.ibm.com/quantum-computing/} or Amazon Braket\footnote{https://aws.amazon.com/braket/} allow them to use the still very expensive to own and operate quantum computers with moderated costs. This model also fits very well with the needs of hybrid systems where both classical and quantum software will be executed on hardware on the cloud regardless of the type of computer needed.

Such deployment architectures are also perfectly aligned with the microservices architectural pattern \cite{dragoni2017microservices}. Following this architecture, a complex systems is conceived as a set of distributed microservices, where each microservice is a cohesive, independent process that interact with the rest of the system through messages. Bringing microservices to hybrid classical-quantum systems, we have solutions where both classical and quantum services coexist to solve complex problems.

To achieve these hybrid microservices architectures, the first step is to convert a quantum piece of software into a microservice that can be integrated with the rest of the architecture. Conceptually, there is no difference between a classical and a quantum microservice, an independent process that can interact with the rest of the system through messages. Furthermore, from a service engineering point of view the specific hardware in which a microservice is executed, classical or quantum, should be irrelevant. However, the current state of quantum services is very different to classical services and requires specific approaches to create working hybrid microservices architectures.  

Running a quantum algorithm as a microservice is possible with the existing technology. The quantum algorithm can be wrapped by a classical service, by using some of the existing quantum software development kits, an integrated into a complex architecture. However, the current technological state of the available quantum platforms impose some limitations. First, they make quantum services invocation and execution to be closely coupled with the quantum processor in which they will be executed. Also, the different quantum computers provide the results of the executed process in different ways thus making even stronger the coupling between the invoking service and the hardware architecture in which the invoked services are executed. In addition, due to the existing noise in quantum computers results are subject to errors and this error are also usually dependent on each specific quantum computer and qbit topology. This increases, again, the coupling between service and hardware. Finally, due to the quantum system collapse, is not always possible to obtain intermediate verification of results which drastically reduces services orchestration possibilities.
 
For all these reasons, invoking a quantum microservice in an agnostic way is not possible and violates all the principles of software engineering. These limitations means that most of the advantages of service-oriented computing are lost when involving quantum microservices. Specially, those related with the different software quality x-abilities like composability, maintainability, reusability, modularity, etc. To address this situation, techniques and methodologies of classical service engineering should be brought to the domain of quantum service engineering \cite{talavera}.   

Specifically, in this paper we focus on the technical aspects needed to create a quantum microservice that can be integrated into a hybrid solution. To illustrate the current state of technology, we have developed a quantum microservice to solve the well known Traveling Salesman Problem (TSP). We use the Amazon Braket platform to deploy this quantum microservice and to analize its characteristics in the different quantum hardware supported by Amazon. Although an specific algorithm (TSP) and platform (Amazon Braket) are used, we believe that the problems and limitations found are directly transferable to other quantum algorithms and platforms. From the results obtained from executing this microservice, we discuss the current limitations of hybrid microservices architectures and detect which of the benefits of service computing are lost in the process, measuring the distance between the current state of technology and the state that would be desirable in order to have a real quantum service engineering.

In order to do that, the rest of the paper is organized as follows. Section \ref{background} present the background of this work. Section \ref{casestudy} details the traveling salesman problem used as the case study for this work, paying special attention to the quantum implementations of this well-known problem. Section \ref{quantumms} presents how to offer a quantum algorithm as a microservice using the Amazan Braket platform. Section \ref{assessment} lists the main results obtained by executing the quantum microservices in the different hardware supported by Braket and discuss the current limitations of quantum microservices. Section \ref{related} address the most relevant related works.  And finally, Section \ref{conclusion} present the paper conclusion and future works.

\section{Background} \label{background}

Microservices in particular, or Service-Oriented Architecture in general, are a software engineering approach focused on the use of services as the fundamental element to develop software solutions \cite{zimmermann2017microservices}. Although different definitions and proposals can be found in the literature, there are some aspects of microservices that are mostly agreed upon and that have led them to be one of the most used paradigm on the cloud.

First, a microservice can be defined as a single-responsibility entity that encapsulates data and logic. They are exposed remotely and can be deployed, changed, substituted, and scaled independently of each other \cite{zimmermann2017microservices}.

When developing microservices solutions, different computing paradigms and storage paradigms can be used \cite{bogner2016towards}. It is common to find microservices solutions in which different programming languages are used, including a mix of functional and imperative ones, and databases, including relational and NoSQL ones, to provide solutions to complex problems.

There is no standard communication mechanism to be used for the development of microservices. Nevertheless, in practice, REST HTTP and asynchronous message queues are the most commonly used ways to expose microservices \cite{bogner2019microservices}.

Similarly, although there is no constraint on where and how microservices should be deployed, in practice most solutions are developed with a strong orientention towards the cloud \cite{sill2016design}. The elasticity and distribution provided by the cloud, alongside the automated management and the loose coupling provided by cloud vendors are features well aligned with the microservices approach.   

Finally, although they are completely independent paradigms that are not exclusively related with microservices, continuous delivery and DevOps approaches are usually applied during the development of microservices systems \cite{chen2018microservices}. The architectural characteristics of microservices mitigates some of the key barriers to adopt these approaches by improving deployability and modifiability. 

Having all this into account, we can assume that microservices will be a good fit for an hybrid classical quantum solution. In this regard, some of the most recent works on quantum software development are helping to align both worlds. Specially, from a cloud computing perspective.

Currently, most commercial quantum computers are accessible through the cloud, similarly to the classical Infrastructure as a Service model. Some researchers have called this access Quantum Computing as a Service (QCaaS) \cite{rahaman2015review}. Through QCaas developers can use some of the existing quantum computers to execute their own code. However, this access is still very dependant of the specific hardware in which the software will be executed, requiring developers to have a deep understanding of the underlying hardware.  

To address some of the QCaaS limitations and increase its abstraction level to create more complex quantum software, multiple research and commercial efforts are underway. 

In the academic world, quantum software engineering is starting to emerge, attracting the attention of researchers \cite{Zhao,Piattini20}. This discipline seeks to bring the knowledge and expertise of classical software engineering to the domain of quantum software development. Specifically, some works are starting to pay attention to aspects of quantum development more closely related with microservices. For example, in \cite{BarzenLFVWW20} the authors propose the term Quantum application as a Service (QaaS) to narrow the gap between classical service engineering and quantum software. Also, \cite{WildBHLVZ20} propose an extension to TOSCA, a standard for software deployment on the cloud, to allow the deployment of quantum services. 

Companies are also trying to help create more complex quantum solutions. Amazon, one of the global leaders in the cloud and computing services domains, has created the above mentioned Amazon Braket. This platform provides a development environment for quantum software engineers.  

At the moment of writing this article, Braket supports hardware from three different vendors (D-Wave, IonQ and Rigetti). D-Wave machines fall on the category of adiabatic quantum computers, while IonQ and Rigetti machines can be classified as circuit-based quantum computing. Having this two different computational models available increases the tools at the disposal of quantum software developers, but also increase the complexity of programming quantum services. To use adiabatic based machines developers must reformulate the problem they want to solve as a quantum annealing metaheuristic specification \cite{boixo2013experimental}. To use circuit based machines developers should know the details of quantum gates and how to adapt the problem they want to solve to a quantum circuit \cite{wille2019ibm}. This makes it more complicated for developers to create independent, maintainable, agnostic quantum microservices.    

Other companies are creating similar platforms, like the above mentioned IBM Quantum or like QPath\footnote{https://www.quantumpath.es/}. However, as far as the authors know, there is no proposal that have addressed the problems and limitations of creating quantum microservices for the development of hybrid solutions.

\section{Traveling Salesman Problem} \label{casestudy}

To illustrate the technical limitations of current quantum platforms for the development of quantum microservices we have chosen a very well known and studied problem. The traveling salesman problem is a famous example in the class of NP-Class problems \cite{warren2013adapting}. It can be categorized as an optimization problem, in which the traveling salesman must visit all cities inside a route, minimizing the traveled distance. Thus, in the classical definition of the problem there exist cities, usually described as nodes, roads connecting those cities, that can be considered as links between these nodes each with a weight indicating the distance. The main drawback of these kind of problems lies in the increasing of possible solutions with the increase of the problem size, i.e. with 5 cities there exist 12 possible routes whereas for 25 cities the number of routes grows to $3.1\times10^{23}$. Furthermore, this particular problem has been expanded into more realistic and complex formulations, usually in the forms of restrictions, such as the case of the (Capacitated) Vehicle Routing Problem \cite{irie2019quantum} or the case of TSP with Time Windows \cite{papalitsas2019qubo}. 

Resolving this problem by classical computing methods it is not always optimal. These methods have been developed for years as replacement to brute force solutions on these optimization problems, but still have certain limitations. In recent years, due to the expansion of quantum computing, researchers have begun to develop quantum algorithms that solve these problems: both for the perspective of adiabatic quantum computing \cite{Warren2020} and for the perspective of gate-based quantum computing \cite{Matsuo2020,Srinivasan2018}.

\subsection{Formulation of the TSP problem}

To formulate the problem we have followed the previous works \cite{ohlmann2007compressed,papalitsas2019qubo}:

\textbf{Definition 1.} \emph{Let $G=(N,A)$ be a directed graph, where $N={0,1,...,n}$ is the finite set of nodes, also known as cities, and $A=N \times N$ is the set of roads or arcs connecting the cities. For every pair $(u,v)$ of cities there exists a road in $A$. A tour is defined by the order in which the cities are visited.}

Typically, in this problem some assumptions are taken in order to facilitate its formulation:

\textbf{Definition 2.} \emph{Let city 0 denote the depot and assume that every tour begins and ends at the depot. Each of the remaining n cities appears exactly once in the tour. We denote a tour as an ordered list $P = (p_0,p_1,...,p_n,p_{n+1})$, where $p_i$ is the index of the city in the i-th position of the tour. According to our previous assumption $p_0=p_{n+1}=0$, which can also be indicated as $i (mod n+1)$.}

\textbf{Definition 3.} \emph{For every pair $(u,v)$ of cities $u,v \in N$, there is a cost $c_{u,v}$, for traversing the road (u, v). This cost of traversing the road from u to v generally consists of the travel time from city u to city v.}

Based on the previous definitions, in the classical formulation of the TSP, the objective function is to minimize the sum of the arc traversal costs along the tour, and can be summarized as:

\begin{equation}
min \sum_{i=1}^{n+1} c_{p-1,p}
\label{eq:TSP}
\end{equation}

In eq. (\ref{eq:TSP}), it is assumed that $(p_0,p_1,...,p_n,p_{n+1})$ is a feasible tour.

\subsection{TSP on Quantum Computing Architectures}

The classical formulation of the problem is proposed as an optimization, making it extremely suitable and straightforward for adiabatic quantum computers. More challenging is the proposal of solutions tailored for quantum computers based on quantum circuits and gates. Nonetheless, there exist in the literature various proposals of the later problem. In this subsection, a solution for the TSP on each type of quantum computers is described.

\subsubsection{TSP on Adiabatic Quantum Computing} \label{sec:tsp_adiabatic}

For this type of quantum computers, Amazon Braket provides a proposal of solution based on quantum annealing. This proposal employs the Lagrange multipliers and Quadratic Unconstrained Binary Optimization (QUBO) problem matrix \cite{Date2019} where the graph of the problem is encoded such that the evolution of the system through quantum annealing offers the minimum energy cost, that is to say, the minimum travel cost.

\begin{figure}[h]
\centering
    \includegraphics[width=0.5\textwidth]{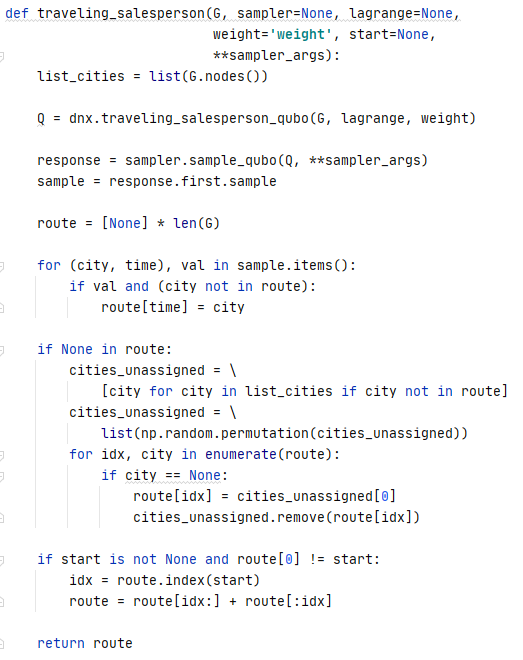}
 	\caption{Fragment of code for the TSP using quantum annealing}
 	\label{fig:traveling_salesperson}
\end{figure}

To do this, the graph is internally converted to the form of an Ising model or Quadratic Unconstrained Binary Optimization Problem (QUBO), and later on quantum annealing is applied, as shown in Figure \ref{fig:traveling_salesperson}.

\begin{figure}[h]
\begin{subfigure}{.5\textwidth}
  \centering
  \includegraphics[width=0.5\linewidth]{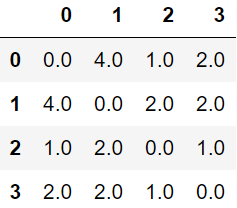}  
  \caption{Input costs' matrix}
  \label{fig:input_annealing_matrix}
\end{subfigure}
\newline
\begin{subfigure}{.5\textwidth}
  \centering
  \includegraphics[width=0.8\linewidth]{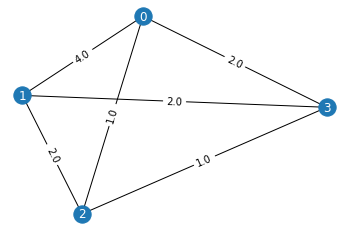}  
  \caption{Generated cities graph}
  \label{fig:input_annealing_graph}
\end{subfigure}
\caption{Input for the annealing-based solution of the TSP}
\label{fig:input_annealing}
\end{figure}

As input for the algorithm, it is necessary to provide a matrix with the costs of traveling between cities. An example of this matrix is showed in Figure \ref{fig:input_annealing_matrix}. From it, a graph like the one on Figure  \ref{fig:input_annealing_graph} can be generated. If the graph is directed, the weights of links connecting 2 nodes will be different, which is the more general representation of the problem, with $(N-1)!$ possible solutions. On the other hand, if a non-directed graph is considered, the weights between 2 nodes will have the same value (such as indicated in Figure \ref{fig:input_annealing_matrix}), having, thus, $(N-1)!/2$ possibles routes. Also, one must note that the starting point is not significant, being route $0\rightarrow 1 \rightarrow 2 \rightarrow 3$  the same as $2 \rightarrow 3 \rightarrow 0 \rightarrow 1$.

\begin{figure}[h]
\centering
    \includegraphics[width=0.4\textwidth]{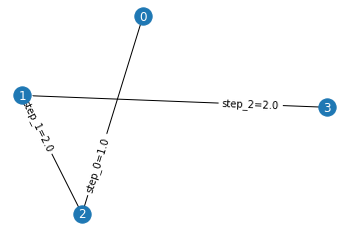}
 	\caption{Output of the annealing-based solution of the TSP}
 	\label{fig:output_annealing}
\end{figure}

As output, the Hamiltonian cycle with the lowest cost is returned \cite{Mahasinghe2019} with 2 considerations: including revisiting the starting point or not revisiting it, i.e., in Figure \ref{fig:output_annealing} it is provided a graphic representation of the path with less cost, without returning to initial point.

\subsubsection{TSP on Gate-based Quantum Computing} \label{sec:tsp_gate}

Solving the TSP in quantum machines programmed via quantum gate-based circuits is much more complicated. Optimization problems, such as TSP, can be addressed simply in adiabatic quantum computers by their nature, but to work on gate-based quantum machines a workaround is needed. This is related to the complexity of defining a circuit that allows to find the solution to the problem in any situation, obtaining in many cases circuits that only solve the problem for a specific case. 
There are different approaches in the literature to solve the TSP in gate-based quantum computers, by means of the application of different quantum algorithms and solutions. From all of the available proposals, in this work the circuit proposed by \cite{Srinivasan2018} will be used. It is based on the Quantum Phase Estimator (QPE) algorithm, which calculates a phase for each of the eigenstates considered. These eigenstates correspond to each of the possible Hamiltonian cycles solutions of the problem. Having obtained the phase of each eigenstate, they are later checked to select the lowest using other quantum algorithms such as the minimum finding \cite{Durr1996}. Thus, the optimal path is the one for which the QPE obtains a minimum phase.

\begin{figure}[h]
\centering
    \includegraphics[width=0.5\textwidth]{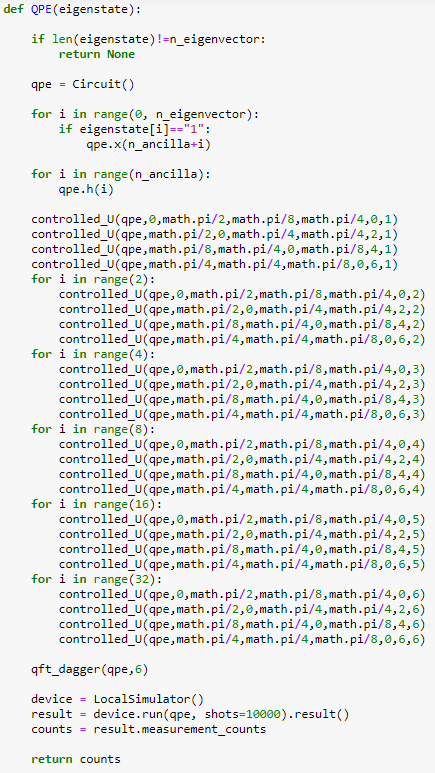}
 	\caption{Gate-based QPE algorithm to solve the TSP}
 	\label{fig:gate_solution}
\end{figure}

The above mentioned quantum algorithm has been implemented in Amazon Braket. A fragment of the resulting code can be seen in Figure \ref{fig:gate_solution}. Starting from the state $\ket{00000000}$, it adds and X gate to the qbits in the positions where it should be 1, thus, for each eigenstate one circuit must be generated; after that, controlled-U gates are applied; finally, a QFT dagger is applied and the results are measured. These controlled-U gates are defined as indicated in \cite{Srinivasan2018}, and each U is an unitary matrix constructed as the tensor product of other Unitaries, i.e. $U=U_1 \otimes U_2 \otimes ... \otimes U_N $. Each of this $U_j$, are obtained after the codification of the initial graph describing the particular values of the problem into a distance/cost matrix such as that for each distance between two cities $i,j$ it is expressed in the matrix as $e^{i·\phi_{ij}}$. This is done after normalizing the original cost matrix from 0 to $2\pi$ range.  

As opposed to the quantum annealing solution, the circuit obtained is not generic, one circuit must be generated for each eigenstate and an implemented circuit is associated with a concrete graph due to the nature of the controlled-U gates and the Unitary matrices obtained. The included code fragment is tailored for the graph shown in Figure \ref{fig:input_annealing}. This way, subsequent experiments and the obtained results will be easier to compare. Nonetheless, one must have in mind that if any of the elements of the graph change, the circuit should be changed as well.

After obtaining the result of the quantum part of the algorithm, classical computations must be applied to determine which eigenstate has as result the phase of smaller value and, knowing which is that eigenstate, return the associated Hamiltonian cycle.

Analyzing both solutions to the TSP, it is clear that the main complication lies in the case of quantum gate-based machines. Making generic circuits, which enable to take as input a series of parameters that condition them ---in this case, a circuit for the QPE algorithm that works with any graph---, is something that has already been studied in proposal like \cite{Matsuo2020,Adelomou2020} that use Parameterized Quantum Circuits. Specifically, Matsuo et al. \cite{Matsuo2020} propose a circuit to solve the TSP for any input graph in gate-based quantum machines, using the VQE algorithm. In this work solutions of this type have not been used since it has not been considered essential for the performed experiments. Being able to solve a given TSP is enough to build a microservice and analyze it.

\section{Quantum microservices} \label{quantumms}

\begin{figure}[h]
\centering
    \includegraphics[width=0.5\textwidth]{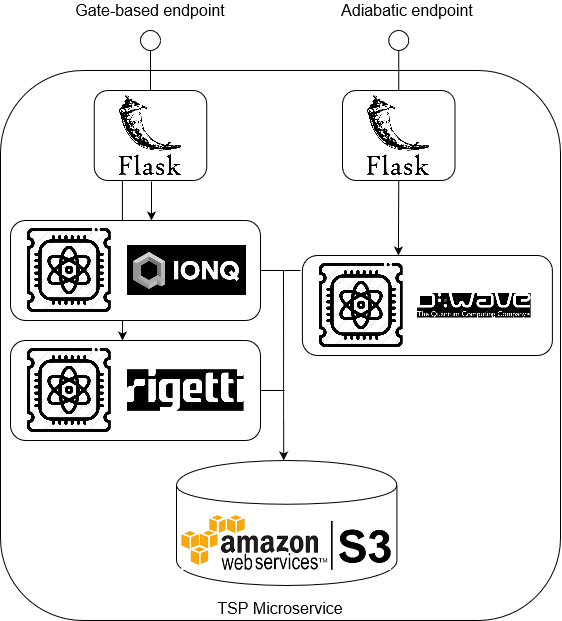}
 	\caption{Deployment architecture of the TSP hybrid microservice}
 	\label{fig:tsp}
\end{figure}

From the two solutions to the TSP discussed in the previous section, a microservice with two endpoints has been implemented as shown in Figure \ref{fig:tsp}. The implementation of this microservice and the notebooks with the implementations of the TSP in adiabatic and gate-based quantum computing are available in the following repository\footnote{https://github.com/frojomar/ICWS2021-quantum-classical-microservices}.

The two endpoints allows other microservices of the system to invoke the solution of the TSP problem, one using the gate-based solution and the other the adiabatic solution. Both endpoints are implemented in Python, using Flask\footnote{https://flask.palletsprojects.com/en/1.1.x/}. This endpoint are in charge of deploying and executing the corresponding quantum algorithm. 

These algorithm are executed on Amazon Braket and, therefore, different quantum computers can be chosen. In the case of the algorithm for adiabatic quantum computing, Braket supports the execution on the quantum computers \textit{D-Wave 2000Q} and \textit{D-Wave Advantage\_system}. In the case of gate-based quantum computing, it can be executed on the simulators \textit{LocalSimulator}, \textit{TN1}, and \textit{SV1} or on the quantum computers \textit{IonQ}, \textit{Rigetti Aspen-8}, and \textit{Rigetti Aspen-9}. In any case, the result of the execution of the quantum algorithms are always stores in an Amazon S3 storage ---except in the case of \textit{LocalSimulator}---.

In order to choose over which hardware the microservice's quantum part is going to be executed, both endpoints employ a parameter. This parameter is codified as a \textit{query param} called \textit{device}. Taking into account the number of QPUs and simulators availables, the query param \textit{device} must take one of the following values:

\begin{itemize}
    \item Adiabatic quantum computing endpoint: \textit{dwave\_dw2000}, and \textit{dwave\_advantage}.
    \item Gate-based quantum computing endpoint: \textit{local}, \textit{tn1}, \textit{sv1}, \textit{ionq}, \textit{riggeti\_aspen8}, and \textit{riggeti\_aspen9}.
\end{itemize}

Next, implementation details of both endpoints will be provided showing that, despite the fact that each one of them belongs to a different type of quantum computing, the way of enclosing the quantum code with a classical computing wrapper, that allows its execution as a microservice, does not differ.

\subsection{Adiabatic quantum computing endpoint} 

\begin{figure}[h]
\centering
    \includegraphics[width=0.5\textwidth]{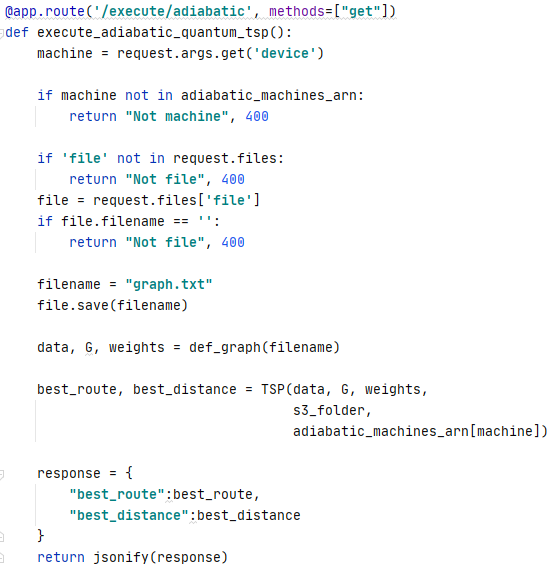}
 	\caption{Adiabatic quantum computing endpoint code.}
 	\label{fig:adiabatic_microservice_code}
\end{figure}

Figure \ref{fig:adiabatic_microservice_code} shows the code for the endpoint of the adiabatic solution.

As input, the endpoint needs a .txt file with the weights of the matrix, in addition to the query param \textit{device}. As a result, a JSON object is returned including the best route found for the TSP problem and the distance to cover the route.


\subsection{Gate-based quantum computing endpoint} 

\begin{figure}[h]
\centering
    \includegraphics[width=0.5\textwidth]{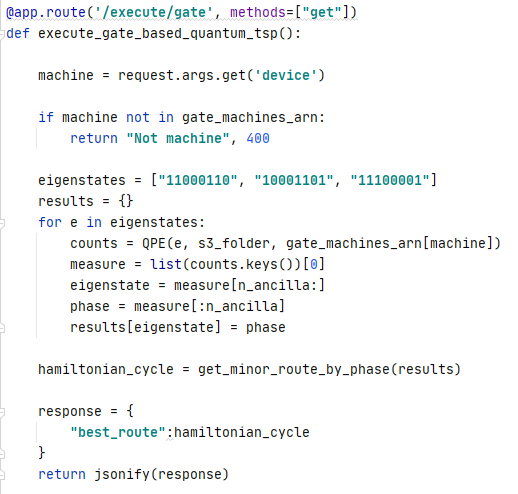}
 	\caption{Gate-based quantum computing endpoint code.}
 	\label{fig:gate_based_microservice_code}
\end{figure}

As in the previous endpoint, Figure \ref{fig:gate_based_microservice_code} shows the code for the endpoint of the circuit based solution. 

Depending on whether the TSP is to be executed on a real quantum computer or on a simulator, the way in which the quantum code is executed changes. Therefore the selected machine is given as a parameter to the method that makes the quantum call. When the endpoint is run outside of a local simulator the result is stored in s3 storage, from where it is retrieved. In particular, if it is run on a quantum computer, the results are always stored in an s3 and, in addition, they take some undefined time to be available.

When running the code on a quantum computer, a task with an identifier is created. With this identifier developers can check the status of the task, which will change from CREATED, QUEUED, and RUNNING, until it reaches COMPLETED; or CANCELLED or FAILED if something goes wrong. At this point, the result can be recovered. In any case, it is necessary to define a poll timeout to prevent the execution from being blocked.

All this is done transparently to the system since, when the microservice is invoked, it is checked with which device the code is to be executed and, if it is a real quantum computer, the wait is done internally in the call to the endpoint.

In order to execute this endpoint, it is necessary to send the query param \textit{device} with the device where the quantum algorithm is to be executed. In this case the quantum circuit is not parameterized, so it does not allow the execution on different networks. That same graph will be always used as mentioned above. As output this endpoint returns the optimal path but not the path cost, as the algorithm does not know the weight of each path.

\section{Quantum-Classical Hybrid Microservices System Trial Evaluation} \label{assessment}

After developing the described microservice, different metrics were used to evaluate its performance and the limitations of including a quantum microservice in a hybrid system.

In order to proceed with the evaluation, several HTTP request were made from the API client tool Postman\footnote{https://www.postman.com/product/api-client/}, that allows performing petitions to REST APIs and take metrics such as response time, response size, etc.

The developed microservice was locally deployed during the evaluation. More specifically, it was deployed on a laptop running Windows 10 with 16 GB of RAM and an Intel Core i7-8550U processor at 1.8 GHz base frequency and 4.0 GHz turbo frequency, equipped with NVMe SSD technology for storage. 

Taking advantage of the fact that the adiabatic solution implemented for the TSP is generic and the endpoint allows giving a graph as input, the evaluation for both adiabatic and gate-based implementations has been carried out using the same graphs as input. Thus, the comparison between the results obtained with both computation models are comparable.

Table \ref{tab:shots} summarizes the launched executions and the corresponding results obtained. The first 2 lines correspond to the Adiabatic Quantum Endpoint and the rest of the table is related to the Gate-based Quantum Endpoint. 

\begin{table}
\centering
 \begin{tabular}{||c c c||} 
 \hline
 Architecture & \# of shots & Result obtained  \\ [0.5ex] 
 \hline\hline
 DWAVE 2000Q6 & $10^2,10^3,10^4$ & [0,3,1,2] (Consistent) \\ 
 \hline
 DWAVE ADVANTAGE & $10^2,10^3,10^4$ & [0,3,1,2] (Consistent) \\ 
 \hline
 \hline
 LocalSimulator & $10^3,10^4,,10^5$ & [0,3,1,2], [0,1,2,3] \\
 \hline
 TN1 & --- & Error \\
 \hline
 SV1 & $10^3,10^4$ & [0,1,2,3] (Inconsistent)  \\
 \hline
 IonQ & --- & Error \\
 \hline
 Aspen 8 & --- & Error \\
 \hline
 Aspen 9 & $10^3$ & Error \\
 \hline
\end{tabular}
\caption{\label{tab:shots} Executions on each endpoint and shots conducted. }
\end{table}

\textbf{Number of qubits}. One of the main limitations of current quantum computers lies in the number of qubits available, specially in the case of gate-based systems, and this directly affect the ability to run the microservice or limit its execution. Table \ref{tab:qubits} shows the results from considering the TSP described in Figure \ref{fig:input_annealing}. As one can see, in the case of Gate-based circuits the number of qubits amounts to 14 (8 for eigenstates + 6 for phase). In the case of the quantum annealing solution the number of qubits necessary is unknown, since the Braket provided implementation was used. Nevertheless, both D-wave machines available at Braket had enough qbits to run the algorithm. In any case, even in for such a simple problem (3 possible routes considering the links between 2 nodes as symmetric non-directed) it exceeds the number of qubits available (11 qubits) to be executed on the IonQ hardware and it is not possible to execute on this architecture. In the case of the Rigetti hardware, it provides enough qubits.

In the case of quantum services, this not only shows the limited power of the current hardware but also the need that quantum service engineering will have for mechanisms to determine the number of qubits the execution of a service will need. Due to the nature of quantum algorithms for the different architectures, there is no trivial way obtain this number. This will be a key question in developing quantum services execution planners with implications in several other aspects of the service, like if cost if only the hardware with more qubits can be used or response time if the waiting time for that hardware is longer.

\begin{table}
\centering
 \begin{tabular}{||c c c||} 
 \hline
 Version & \# of Qubits & \# of Classical bits  \\ [0.5ex] 
 \hline\hline
 Gate-based TSP & 14 (eigenstates+ phase) & 6 (collapsing phase) \\ 
 \hline
 Dwave's solution & Unknown & Unknown \\
 \hline
\end{tabular}
\caption{\label{tab:qubits} Executions conducted and number of qubits and classical bits}
\end{table}

\textbf{Number of shots} Due to the problems that arise due to the characteristics of actual quantum computers, mainly noise in the qubits state, the experiments must be conducted several times or "shots" to be statistically consistent. 

For a real quantum service technology, the responsibility of performing the different executions to get a consistent result cannot be delegated in the client nor the customer who only wants to use a technology to get a correct result, at least within a given margin of error, and with an economic cost known in advance. How the number of shots required is estimated will have a direct impact in the cost of the service executions. This reveal some issues, related with service quality and costs, that still have to be addressed by quantum services engineering.

\textbf{Precision of results} Table \ref{tab:shots} shows the discrepancies in the results achieved. In the first rows, the results obtained by DWave's machines are shown. Both show consistent results given the number of shots considered. For the rest of platforms different problems arisen. First, in some of the architectures it has been impossible to execute the code, more specifically it happened in TN1, IonQ and Rigetti 8. On IonQ the number of qubits available was insufficient to run the service. On Apen 8, the service was unavailable at the time of running the experiments. Finally, in Aspen 9, although the code was send for execution, it was in the state QUEUED for more than 3 hours, not having executed any of the 1000 shots. In contrast, in the case of SV1, the code was sent and executed. However, the result obtained were different in different executions. Most of the times the result was [0,1,2,3] which does not correspond to the optimal solution. Similar results were observed when working with the simulated architecture, where the number of shots must be higher than 100 in order to obtain the correct solution with acceptable statistical certainty.

Again, in a real scenario, the responsibility of determining the number of shots and the precision of the results cannot be delegated in the client service nor the customer. The customer pays for a service that is expected to provide correct results, within an agreed level. This point outs to the needs of some kind of logic in service execution planners to determine the number of shots needed to provide a correct solution. It should also be noted that predictions can affect the planning, availability and accuracy of the platforms' results and all of this will impact on the service qualities that will have been previously negotiated with the customers. These are therefore issues that, while affecting the technical aspects of future quantum services platforms, will also affect their financial profitability.

\textbf{Response times} Other evaluated parameter was response time. The measure corresponds to the time elapsed between sending the request and receiving the result. This time has been measured for all the machines where the code has been correctly executed. Specifically, the SV1 and LocalSimulator gate-based simulators, and the D-wave 2000\_Q6 and D-wave Advantage adiabatic quantum machines. For the first ones, the difference is significant. In the LocalSimulator, the execution took about 3 seconds with 1000 and 10000 shots, and about 7 seconds with 100000 shots. However, in SV1 it took an average of 27 seconds, with a margin of up to 10 seconds between the fastest and slowest runs. In the case of adiabatic machines, the result is similar for both. In the case of D-wave 2000\_Q6, the runs exceed 20 seconds and in the case of D-wave Advantage 25 seconds. 

From this, it can be concluded that the highest cost in terms of time is incurred when sending the quantum code to execution. Possibly, due to the waiting times in the queue. Nevertheless, these results gives the user a feeling of unreliability when using the platforms which, in real service engineering, must be avoided. Dealing with this will again require quantum service platform planners to count on reliable resources and estimates. This highlights how far current quantum service platforms from reaching to be acceptable to potential customers of a quantum services platform. 

\textbf{Economic cost} Lastly, the economic cost of invoking each solution has been considered. At the moment, Amazon Braket establish a fixed price per quantum task executed, that is the same for all the supported hardware. Moreover, an additional cost is paid for each shot and this cost is different for the different hardware. These costs, while predictable if the hardware to be used and the number of shots to be executed are known, are far from what is needed to agnostically implement microservices on hybrid quantum-classic systems. Specially due to the uncertainty that arises from unavailability of services, response time, uptime, state of the quantum system and so on, parameters extremely important to be able to assure the quality and SLAs inherent of services. 

Furthermore, from the evaluation performed, some other more abstract questions arise as well. First and foremost, there is a need for abstractions to define quantum problems in a more general way. This abstraction could be used as an starting point that can be specialized in terms of quantum annealing, gate-based circuits or whatever future technology or new programming paradigm appear for quantum computing. An initial solution, for the specific case presented in this paper could be to develop a single generic endpoint in charge of unifying the adiabatic and gate-based solutions. Such an endpoint will have the responsibility of adapting an abstract representation of the TSP problem to the needs of the specific quantum hardware in which it will be executed. Even then, the solution will still depend on the service platform, Braket in this case. This remarks one unresolved question in quantum service engineering. When a service is invoked, the invoker only cares for the response. The service platform should address the execution on different architectures and the problem formulation for each of them if there is a benefit on doing so. Delegating these responsibilities on the client makes it more tightly coupled with the microservice and reduce the benefits provided by service engineering. 

Summarizing, given the above mentioned problems, current quantum services platforms pose the following inconveniences for the development of quantum microservices. First, services are tightly coupled with the quantum code to be executed. Moreover, services are also tightly coupled with the hardware in which the quantum code will be executed. Additionally, platforms do not allow a service implementation to be transparently replaced by another, as can happen in traditional services as long as the API is maintained. Also, quantum platforms are not able to decide, on execution time, where and how a service will be executed to optimize answering petitions based on performance aspects of the different supported hardware. Finally, all the experiments developed in this paper involve only a single service which is completely unreal. The most simple example of a real microservices based system would involve several ones. However, there is also an absolute lack of mechanisms for quantum services orchestration. All of these limitations have a significant impact in some of the most relevant aspect of quality services, like composability, maintainability, reusability or modularity of quantum services limiting the current potential of these services. These limitations affect not only researchers or developers but also the platforms that provide access to quantum hardware. The commercial success of cloud computing and services is supported by the elasticity provided to developers and the optimization of hardware usage provided to the hardware owners. Similar levels of flexibility and optimization should be possible in the quantum domain but additional research efforts are needed in quantum service engineering.

\section{Related works} \label{related}

Quantum software engineering is a young discipline and works that focus on quantum microservices or hybrid microservices architectures are still sparse. However, some researchers are starting to focus on this and related topics.

Works like \cite{Leymann2020QuantumCloud} start to explore the potential of quantum services in the cloud and the research opportunities of quantum services. Some of the research opportunities identified in the work remark the problems found in this paper to develop quantum microservices. in particular, it includes the problems caused by needing different implementations of the same quantum algorithms for different hardware vendors or the problems to deploy quantum services in quantum computers. 

As mentioned before, in \cite{WildBHLVZ20} the authors further explore the deployment of quantum services by proposing a TOSCA extension to deploy quantum software. This proposal share some similarities with the work presented here. Since quantum applications must be  newly deployed for each invocation, a classical computer is needed to host and deploy them. Here we have use a classical web service to wrap the quantum algorithms and to expose them as endpoints.

Similarly, in \cite{sim2018framework} the authors propose a technique for algorithm deployment on cloud based quantum computers. The proposed technique is only valid for circuit based quantum algorithms as it start from a generic quantum circuit that is later compiled for a specific quantum computer.

To try to minimize some of the vendor lock-in problems caused by quantum algorithms being very closely related to the hardware in which they will be executed, some researchers are also working on parameterized quantum circuits \cite{Matsuo2020,Adelomou2020}. This technique allows to create circuits that can be modified based on input parameters which can be used to adapt them to different hardware or problems. However, this technique can not be applied to quantum annealing based hardware.

From a commercial perspective, along with Amazon Braket, there are other proposals related to the simplification and homogenization of quantum access to machines and services. Such is the case of Azure Quantum \cite{cuomo2020towards}, an alternative to Amazon Braket. Azure Quantum provides a development kit for quantum software that tries to unify an heterogeneous set of hardware and software solutions.

Other companies and software developers are creating high level development environments, toolkits and APIs to increase the abstraction level of quantum software. For example IBM proposes IBM Quantum\cite{cross2018ibm} while other focus on specific domains like quantum machine learning\cite{hu2019quantum}. However, as far as the authors know they do not provide any specific advantage for the development of quantum microservices or hybrid solutions over Amazon Braket.

Moreover, to develop quantum microservices with a similar quality  that classical services it is not enough to simplify the development and deployment of quantum algorithms. Other aspects of service engineering\cite{LiZJZZSSB21} can not be overlooked. 

Some researchers are focusing on the orchestration aspects of quantum complex algorithms. In \cite{cohen2020quantum} the authors propose a hardware based orchestrator to control the flow of complex quantum and hybrid applications. However, for quantum microservices to be used with the same ease that classical services, software orchestration solutions are still needed. 

In a similar way, once quantum microservices are deployed and orchestrated they have to be operated. To do that, quantum DevOps is starting to emerge. In \cite{gheorghe2020quantum} the author propose a technique to regularly check the reliability of quantum computers. This reliability is the used to estimate if a given hardware will provide results of enough quality and to select the best fitting available hardware to run a quantum service.

Finally, in \cite{phalak2021quantum} the authors focus on trust and security issues in quantum services. The current model in which quantum services are managed introduces some trust issues regarding the specific hardware in which a given quantum task is run and other related issues.

All these works reveal that, as identified in this paper, more research is needed in order to develop an effective quantum service engineering discipline.

\section{Conclusion and future works} \label{conclusion}

In this paper we have presented an implementation of a quantum microservice and the problem that arise from trying to integrate it in an hybrid microservices architecture. We have used Amazon Braket to test the implemented microservice on quantum hardware from three different vendors and to detect current limitations in the domain of hybrid classical quantum microservices.

The performed experiments have allows us to clearly show the limitations of current quantum computers platform for the development and exploitation of quantum microservices. Intense research efforts are still needed to bring the benefits of service-oriented computing and microservices to the quantum computing domain.

Since quantum software engineering is still a very young discipline, most of the areas that compose it are just starting to attract the interest of researchers, including hybrid microservices architectures. However, the change in computing paradigm that implies quantum computing means that we cannot directly translate the techniques and tools of classical microservices and expect them to work flawlessly in the new environment. Putting a quantum algorithm inside a microservice is not enough to create a quantum microservices, there needs to be an effort to generate new knowledge, techniques, metodologies,... that helps bridge the gap between classical microservices and the advantages of quantum computers. 

\section*{Acknowledgment}
This work was supported by the projects 0499\_4IE\_PLUS\_4\_E (Interreg V-A España-Portugal 2014-2020) and RTI2018-094591-B-I00 (MCIU/AEI/FEDER, UE), by the FPU19/03965 grant, by the Department of Economy and Infrastructure of the Government of Extremadura (GR18112, IB18030), and by the European Regional Development Fund.

\bibliographystyle{./bibliography/IEEEtran}
\bibliography{./bibliography/IEEEabrv,./bibliography/IEEEexample,./bibliography/biblio.bib}

\end{document}